\documentclass[preprint,aps,floatfix]{revtex4-1}
\usepackage{hyperref}
\everymath={\displaystyle}
\usepackage{young}
\usepackage{rotating}
\usepackage{longtable}
\def\1{\mbox{l\hspace{-0.53em}1}}

\usepackage{multirow}

\newlength{\AccoHaut}

\begin{document}
\title{SU(3) Clebsch-Gordan coefficients at large $N_c$}

\author{Fl. Stancu\footnote{E-mail address: fstancu@ulg.ac.be}}
\affiliation{University of Li\`ege, Institute of Physics B5, Sart Tilman,
B-4000 Li\`ege 1, Belgium}

\date{\today}

\begin{abstract}
It is argued that several papers where SU(3) Clebsch-Gordan coefficients were calculated in order to 
describe properties of hadronic systems are, up to a phase convention,
particular cases of analytic formulae derived  by Hecht in 1965 in the context of nuclear physics.  
This is valid for irreducible representations with multiplicity one in the corresponding 
Clebsch-Gordan series. For multiplicity two, Hecht has proposed an alternative which can provide 
correct $1/N_c$ sub-leading orders  in large $N_c$ studies.	

\end{abstract}

\maketitle

\section{Introduction}

Since the 1963 classical paper of de Swart \cite{de Swart:1963gc}  where Clebsch-Gordan (CG) coefficients 
of SU(3) were derived for the most important direct (or Kronecker) products of irreducible representations
needed in particle physics at that time, namely  
 $\bf 8 \times {\bf 8}$,  $\bf 8 \times {\bf 10}$,  $\bf 8 \times {\bf 27}$,
 ${\bf 10} \times {\bf 10}$ and  ${\bf 10} \times {\bf \bar{10}}$,
many authors devoted their papers or parts of them to the derivation of CG coefficients which
were missing in de Swart's paper. As recalled in the next section this amounts to derive the corresponding 
isoscalar factor for each CG coefficient. 
In 1963 as well, numerical values for the SU(3) isoscalar factors were published by Edmonds \cite{EDMONDS}.
More tables were given in 1964 by McNamee and Chilton \cite{McNamee:1964xq}.

In recent years the SU(3) flavor group was frequently used to study new hadronic 
properties and quark systems involving an arbitrary number of quarks as 
for example in large $N_c$ QCD studies. The existing results seemed to be insufficient so that several
authors derived their own tables. Here we show that some of them are 
particular cases of the analytic expressions obtained by Hecht in 1965 in the context of nuclear physics
\cite{HECHT}.

The purpose of this note is twofold: a) To draw attention to  Hecht's work, which 
may not be known by particle physicists. Some 
analytic formulae obtained by Hecht for SU(3) isoscalar factors  can straightforwardly 
be used for particular cases. 
b) To show that Hecht's results for multiplicity two in the direct products 
are useful in large $N_c$ studies, which give a qualitative insight into the 
structure of baryons. In the large $1/N_c$ expansion one has first to analyze 
formulae at arbitrary $N_c$ and afterwards take $N_c$ = 3 in applications. 

\section{Reminder of some SU(3) CG properties}\label{Reminder}

In the chain $SU(3) \supset SU(2)_I \times U(I)_Y$ each SU(3) CG coefficient factorizes into an
SU(2)-isospin CG coefficient and an SU(3) isoscalar factor \cite{de Swart:1963gc}
\begin{equation}\label{CGSU3}
\left(\begin{array}{cc|c}
	(\lambda \mu)    &  (\lambda_a \mu_a)   &   (\lambda'\mu')\\
	YII_3   &  Y^aI^aI^a_{3}  &  Y' I' I'_3
      \end{array}\right)_{\rho} =
\left(\begin{array}{cc|c}
	I   &    1  &  I'   \\
	I_3 &    I^a_3  &  I'_3
     \end{array}\right)
 \left(\begin{array}{cc||c}
	(\lambda \mu)    &  (\lambda^a \mu^a)   &   (\lambda'\mu')\\
 	 YI   &  Y^aI^a  &  Y' I'
      \end{array}\right)_{\rho}.
 \end{equation} 
where  $(\lambda \mu)$ labels an SU(3) irreducible representation (irrep) and  the index $\rho$
distinguishes between identical representations occurring in the decomposition of a given direct product 
where the multiplicity of $(\lambda'\mu') = (\lambda \mu)$
is larger than one. The highest multiplicity considered here is two
and in this case
a typical example of direct product representations is when one takes $(\lambda^a \mu^a)$ = (11),
which is the adjoint representation of SU(3), also denoted by its dimension {\bf 8}. The CG series reads
\begin{eqnarray}\label{PROD}
\lefteqn{(\lambda \mu) \times (11)  =  (\lambda+1, \mu+1)+ (\lambda+2, \mu-1) +
(\lambda \mu)_1 + (\lambda \mu)_2}  \nonumber \\
& & + \, (\lambda-1, \mu+2) + (\lambda-2, \mu+1)
+ (\lambda+1, \mu-2)+ (\lambda-1, \mu-1).
\end{eqnarray}
The isoscalar factors of SU(3) satisfy an orthogonality relation resulting from the 
orthogonality relations of SU(3) and SU(2) CG coefficients. This is
\begin{widetext}
\begin{equation}\label{ORTHOG}
\sum_{Y'' I'' Y^a I^a}
 \left(\begin{array}{cc||c}
	(\lambda'' \mu'')    &  (\lambda^a \mu^a)   & (\lambda' \mu')   \\
	 Y'' I'' &  Y^a I^a   &   Y I
      \end{array}\right)_{\rho}
   \left(\begin{array}{cc||c}
	 (\lambda'' \mu'')   &  (\lambda^a \mu^a)   & (\lambda \mu) \\
	Y'' I''  &   Y^a I^a  &   Y' I'
      \end{array}\right)_{\rho} = \delta_{\lambda' \lambda} 
      \delta_{\mu' \mu} \delta_{Y' Y}\delta_{I' I},   
 \end{equation}
 \end{widetext}
and
\begin{widetext}
\begin{equation}\label{ORTHOG1}
\sum_{(\lambda \mu) \rho}
 \left(\begin{array}{cc||c}
	(\lambda'' \mu'')    &  (\lambda^a \mu^a)   & (\lambda \mu)   \\
	 Y'' I'' &  Y^a I^a   &   Y I
      \end{array}\right)_{\rho}
   \left(\begin{array}{cc||c}
	 (\lambda'' \mu'')   &  (\lambda^a \mu^a)   & (\lambda \mu) \\
	Y''_1 I''_1  &   Y^a_1 I^a_1  &   Y I
      \end{array}\right)_{\rho} = \delta_{Y''Y''_1}  \delta_{I''I''_1}
      \delta_{Y^a Y^a_1} \delta_{I^a I^a_1}.   
 \end{equation}
\end{widetext}
For completeness, we also recall that the isoscalar factors obey the 
following symmetry properties \cite{HECHT}

\begin{eqnarray}\label{SYM1}
\lefteqn{ \left(\begin{array}{cc||c}  (\lambda \mu)  &  (\lambda^a \mu^a)  &  (\lambda' \mu') \\
                              YI            &  Y^aI^a & Y'I'
                                      \end{array}\right)=} \nonumber \\  & & 
(-)^{(\lambda-\mu+\lambda^a-\mu^a-\lambda'+\mu'+I+I^a-I')}
\left(\begin{array}{cc||c}   

                       (\lambda^a \mu^a)  &  (\lambda \mu)     &  (\lambda' \mu') \\
                         Y^aI^a &      YI &  Y'I' 
                        \end{array}\right).
\end{eqnarray}
and
\begin{eqnarray}\label{SYM2}
\lefteqn{ \left(\begin{array}{cc||c}  (\lambda \mu)  &  (\lambda^a \mu^a)  &  (\lambda' \mu') \\
                              YI            &  Y^aI^a & Y'I'
                                      \end{array}\right)=} \nonumber \\  & & 
(-)^{\frac{1}{3}(\mu'-\mu-\lambda'+\lambda+\frac{3}{2}Y^a)+I'-I}
\sqrt{\frac{\mathrm{dim}(\lambda'\mu')(2I+1)}{\mathrm{dim}(\lambda\mu)(2I'+1)}}
\left(\begin{array}{cc||c}   

                             (\lambda' \mu')  &  (\lambda^a \mu^a)  &  (\lambda \mu) \\
                              Y'I' & -Y^aI^a &  YI 
                                      \end{array}\right).
\end{eqnarray}
where $\mathrm{dim}(\lambda\mu) = \frac{1}{2}(\lambda+1)(\mu+1)(\lambda+\mu+2)$ 
is the dimension of the irrep $(\lambda\mu)$ of SU(3).			      
An alternative notation of the isoscalar factors is $\langle (\lambda\mu)Y I;(\lambda^a\mu^a)Y^a I^a||(\lambda'\mu')Y' I' \rangle$,
see Hecht's paper.

\section{Calculation of SU(3) Clebsch-Gordan coefficients}

The usual procedure to calculate CG coefficients is to start from the highest weight basis vector
of a representation and use ladder operators, which  are $U_{\pm}$, $V_{\pm}$ and $I_{\pm}$ in SU(3).
Their matrix elements were first determined by Biedenharn \cite{Biedenharn1962}. 
Recursion relations among Clebsch-Gordan coefficients are obtained by coupling two states,
as in the usual way, like for the rotation group. These recursion relations contain isoscalar factors.

To  uniquely define the matrix elements of the ladder operators some phase conventions must be made.
For the states in the same isomultiplet the standard Condon and Shortley has been chosen. Accordingly the 
non-vanishing matrix elements of $I_{\pm}$ are positive. 
The relative phases between different 
isomultiplets were defined by the requirement that the non-vanishing matrix elements of $V_{\pm}$ are real and positive
\cite{de Swart:1963gc} (for the phase convention of de Swart  see \cite{de Swart:1963gc}, Section 10).

This procedure has been followed by  Kaeding \cite{Kaeding:1995vq} who provided a large number of tables for 
$(\lambda^a \mu^a)$ = (10), (01), (20), (11), (30) and (21) or in dimensional notation ${\bf 3, \bar{3}, 6, 8, 10}$    
and ${\bf 15'}$. 

More recently Hong  \cite{Hong:2011ch} has derived the isoscalar factors of the direct product of ${\bf 35} \times {\bf 8} $,
with the purpose of using them to the calculation of baryon magnetic moments and decuplet-to-octet transition magnetic moments.
For multiplicity one, all the isoscalar factors  are particular cases of the formulae derived by Hecht 
\cite{HECHT} in his Table 4, up to a phase convention (see next section).

In large $N_c$ QCD Cohen and Lebed \cite{Cohen:2004ki} derived $N_c$ dependent SU(3) CG coefficients 
relevant for the coupling of large $N_c$ baryons to mesons. 
They 
provided extended tables for the direct products for
\begin{equation}
 (\lambda \mu)  = (1, \frac{N_c-1}{2}),\ \ \ (3, \frac{N_c-3}{2}) 
\end{equation}
denoted by ${ "\bf 8"}$ and ${ "{\bf 10}"}$ respectively and $(\lambda^a \mu^a)$ = (11) denoted by ${\bf 8}$. 
Their 
results, at multiplicity one, up to an overall phase, can directly be reproduced    
from Hecht's Table 4.  For
multiplicity two, for example, ${\bf "10"}_a \times {\bf 8} \rightarrow {\bf "10"}_a$  
they are different at arbitrary $N_c$, but 
identical at $N_c$ = 3, as compared to those derived here using Hecht's analytic forms (see next section). 

For the same direct products as those of Cohen and Lebed \cite{Cohen:2004ki}
partial tables were previously provided in Ref. \cite{Diakonov:2013qta}.

The explicit algebraic expressions derived by Hecht \cite{HECHT} for SU(3) isoscalar factors were intended to 
nuclear physics applications, in particular to describing rotational states of deformed light nuclei from the $2s-1d$ shell. 
The deformed nuclei possess collective states described  by Elliott \cite{Elliott:1958zj,Elliott:1958yc} in a model 
where the SU(3) group is used. Thus the application of SU(3) in  nuclear 
physics in 1958 predates the SU(3) classification of elementary particles of
Gell-Mann \cite{GellMann:1962xb} and Ne'eman \cite{Ne'eman:1961cd} 
in 1961. 
The basic reason of using SU(3) in nuclear models is that intrinsic levels of nuclei can be described by the harmonic oscillator 
and SU(3) is the 
symmetry group of the harmonic oscillator in three dimensions (see, for example,
Ref. \cite{Stancu:1991rc} chapter 8).
The physical states of a given angular momentum can be obtained by a projection technique \cite{HECHT1}.

In addition to the isoscalar factors needed for the $2s-1d$ shell, Hecht had also derived explicit expressions for the 
direct product $(\lambda \mu) \times (11)$, considering such results as being of interest,
not surprisingly, because (11) is the adjoint representation of SU(3).
He used the 
standard technique of generating CG coefficients through recursion formulae containing matrix elements of the 
SU(3) generators, but introduced a phase convention different from that of de Swart. The difference
is clearly explained in a footnote of Ref. \cite{HECHT}. In addition, when the 
irrep $(\lambda \mu)$ appears twice in the decomposition of the direct product  
 $(\lambda \mu) \times (11)$, see Eq. (\ref{PROD}),
he introduced the quantum number $\rho$ to label the independent modes
of coupling, such as to have non-zero matrix elements of the SU(3) generators for only one state $\rho$. 
Then, according to the Wigner-Eckart 
theorem, the matrix elements of the generators $T^a$ of SU(3) are
\begin{eqnarray}\label{FLAVOR}
\lefteqn{\langle (\lambda'\mu') Y' I' I'_3; S' S'_3 |T^a|
(\lambda \mu) Y I I_3; S S_3 \rangle =} \nonumber \\ & &
\delta_{SS'} \delta_{S_3S'_3}\delta_{\lambda \lambda'} \delta_{\mu\mu'}
\sum_{\rho = 1,2}
\langle (\lambda'\mu') || T^{(11)} || (\lambda \mu) \rangle_{\rho}
  \left(\begin{array}{cc|c}
	(\lambda \mu)    &  (11)   &   (\lambda'\mu')\\
	YII_3   &  Y^aI^aI^a_{3}  &  Y' I' I'_3
      \end{array}\right)_{\rho},
   \end{eqnarray}
where the reduced matrix elements are defined as  \cite{HECHT} 
\begin{eqnarray}\label{REDUCED}
\langle (\lambda \mu) || T^{(11)} || (\lambda \mu) \rangle_{\rho} = \left\{
\begin{array}{cc}
\sqrt{C(\mathrm{SU(3)})}      & \mathrm{for}\ \rho = 1 \\
0 & \mathrm{for}\ \rho = 2 \\
\end{array}\right.
,\end{eqnarray}   
in terms of the eigenvalue of the Casimir operator       
$C(\mathrm{SU(3)}) = \frac{1}{3} g_{\lambda \mu}$ where
\begin{equation}\label{CSU3}
g_{\lambda\mu}= {\lambda}^2+{\mu}^2+\lambda\mu+3\lambda+3\mu.
\end{equation}
Such a definition is useful for extending the method of calculation of isoscalar factors 
to other SU(N) groups. It has been applied to the calculation of the matrix elements 
of SU(6) generators, where one takes into account that SU(3) is a subgroup of SU(6) \cite{Matagne:2008kb,Matagne:2011fr}.

The correspondence with other notations is 
\begin{eqnarray}\label{rho}
\rho = 1 \Longleftrightarrow {\bf (\lambda\mu)}_2 \Longleftrightarrow {\bf (\lambda\mu)}_a, \nonumber \\ 
\rho = 2 \Longleftrightarrow {\bf (\lambda\mu)}_1 \Longleftrightarrow {\bf (\lambda\mu)}_s.
\end{eqnarray} 
where $s$ and $a$ stand for symmetric and antisymmetric respectively \cite{Lichtenberg,CORNWELL}.
Historically,  following Gell-Mann, in Eq. (\ref{rho}),
it is customary to call the 
symmetric combinations $D$ coupling and the antisymmetric $a$ combinations $F$ coupling
(the $F$ and $D$ notation is used in Ref. \cite{Diakonov:2013qta}, for example).

Ambiguities in distinguishing the  representations at multiplicity larger than one
are typical for all groups, including the permutation
group \cite{ISOSC}. 

Another way to derive Clebsch-Gordan coefficients for SU(3) is based on the tensor
method (for an introduction see, for example, Ref. \cite{Stancu:1991rc}, Sec. 8.10). This method has been used 
for the Clebsch-Gordan series  $\bf "8" \times {\bf 8}$ and  $\bf "10" \times {\bf 8}$ 
in the systematic analysis of large $N_c$ baryons \cite{Dashen:1993jt}.


\section{Examples}

Here we wish to demonstrate the usefulness of Hecht's results, especially for multiplicity two, by using Table 4 of Ref. \cite{HECHT}.
We use the same table format as that of de Swart because it helps in comparing with
previous results found in the literature and moreover, it allows easy checking of the orthogonality 
relations (\ref{ORTHOG}) and (\ref{ORTHOG1}).
We consider two examples relevant for our purpose.

\subsection{Example 1}
The first example, shown in Table \ref{Y2I2}, corresponds to one table obtained by Hong in Ref. \cite{Hong:2011ch}. 
It contains the isoscalar factors  for all irreducible representations with $Y$ = 2, $I$ = 2  
from the decomposition of the direct product ${\bf 35} \times  \bf 8$. These are 
${\bf 81}$, ${\bf 64}$, ${\bf 35}_s$ and ${\bf 35}_a$ in this case.
 
Note that one must use the symmetry property (\ref{SYM1}) to recover the phases for 
$ \bf 8 \times {\bf 35}$ as in Ref. \cite{Hong:2011ch}, because here we consider  ${\bf 35} \times  \bf 8$.
For the columns ${\bf 81}$ and ${\bf 64}$ the absolute values are the same as those of Hong.
Incidentally column ${\bf 81}$ also has  the same phases as Hong and column ${\bf 64}$ has an overall opposite phase.
Our results for ${\bf 35}_s$ and  ${\bf 35}_a$ are entirely different from those of  \cite{Hong:2011ch}
because the definition is different. In applications care must be taken in passing from one convention
to another, especially for calculating transition matrix elements.

\begin{table}
\caption{Isoscalar factors for the irreducible representations with $Y$ = 2, $I$ = 2 
from the decomposition of the direct product ${\bf 35} \times \bf 8$. The first two columns indicate 
the hypercharge and isospin of ${\bf 35}$ and $\bf 8$   respectively. The phase convention is that of Hecht \cite{HECHT}. }
\label{Y2I2}
\[
\renewcommand{\arraystretch}{1.5}
\begin{array}{crrrr}
\hline
 Y_1 I_1 ;\ \  Y_2 I_2  & \ \ \ \ \ \ \ \  ${\bf 81}$  & \ \ \ \ \ \ \ ${\bf 64}$  & \ \ \ \ \ \ \ ${\bf 35}$_s & \ \ \ \ \ \ \ ${\bf 35}$_a \\
\hline
 1,\frac{5}{2};  \ \  1,\frac{1}{2}  & - \sqrt{\frac{1}{200}}   &   \sqrt{\frac{8}{25}} & \sqrt{\frac{5}{8}}  &  \sqrt{\frac{1}{20}} \\
 1,\frac{3}{2};  \ \  1,\frac{1}{2}  &   \sqrt{\frac{144}{200}} &   \sqrt{\frac{2}{25}} &       0                 & -\sqrt{\frac{4}{20}} \\
  2, 2 ; \ \ \   0, 1                &   \sqrt{\frac{10}{200}}  &   \sqrt{\frac{5}{25}} & -\sqrt{\frac{2}{8}}       &  \sqrt{\frac{10}{20}} \\
  2, 2 ; \ \ \   0, 0                &   \sqrt{\frac{45}{200}}  & - \sqrt{\frac{10}{25}}&  \sqrt{ \frac{1}{8}}    &  \sqrt{\frac{5}{20}}  \\[0.5ex]
\hline
\end{array}
\] 
\end{table}

\begin{sidewaystable}
\caption{Isoscalar factors for the irreducible representations with $Y$ = $N_c/3$, $I$ = 3/2 
from the decomposition of the direct product $\bf "10" \times {\bf 8}$ obtained from Table \ref{Hechtp31}.}
\label{CL1}
\[
\renewcommand{\arraystretch}{1.7}
\begin{array}{ccclc}
\hline
 Y_1 I_1; \ \ \ \ Y_2 I_2  & \ \ \ \ \ \  ${\bf "35"}$  & \ \ \ \ ${\bf "27"}$  & \ \ \ \ \ \ \ \ ${\bf "10"}$_a &  ${\bf "10"}$_s\\
\hline
 \frac{N_c}{3},\frac{3}{2}; \ \ \ \ 0,1 & \ \  \sqrt{\frac{12}{16(N_c+9)}}  &  \ \ \sqrt{\frac{5}{4(N_c+1)}} & \ \ \sqrt{\frac{45}{N^2_c+6N_c+45}}
 & -\sqrt{\frac{(N_c-3)(N_c+5)(N_c+6)^2}{(N_c+1)(N_c+9)(N^2_c+6N_c+45)}}\\
 \frac{N_c}{3},\frac{3}{2}; \ \ \ \ 0,0  & \ \ \sqrt{\frac{60}{16(N_c+9)}}   & \ \ - \sqrt{\frac{9}{4(N_c+1)}} & \
 \ \sqrt{\frac{N^2_c}{N^2_c+6N_c+45}}
 & \sqrt{\frac{45(N_c-3)(N_c+5)}{(N_c+1)(N_c+9)(N^2_c+6N_c+45)}}\\
 \frac{N_c}{3}-1,1; \ \  1, \frac{1}{2}   &  \ \ \sqrt{\frac{15(N_c+5)}{16(N_c+9)}}   & \ \ \sqrt{\frac{N_c+5}{16(N_c+1)}} & 
 \ \ - \sqrt{\frac{9(N_c+5)}{4(N^2_c+6N_c+45)}} &  \sqrt{\frac{5(N_c-3)^3}{4(N_c+1)(N_c+9)(N^2_c+6N_c+45)}} \\
 \frac{N_c}{3}-1,2; \ \  1,\frac{1}{2}    &  \ \ - \sqrt{\frac{(N_c-3)}{16(N_c+9)}}   & \ \ \sqrt{\frac{15(N_c-3)}{16(N_c+1)}} &
\ \ \sqrt{\frac{15(N_c-3)}{4(N^2_c+6N_c+45)}} &
 \sqrt{\frac{3(N_c+5)(N_c+21)^2}{4(N_c+1)(N_c+9)(N^2_c+6N_c+45)}}  \\[0.5ex]
\hline
\end{array}
\] 
\end{sidewaystable}

\subsection{Example 2}

The second example is exhibited in Table \ref{CL1} and corresponds to a table of Cohen and Lebed  \cite{Cohen:2004ki},
containing isoscalar factors with $Y$ = $N_c/3$, $I$ = 3/2
from the decomposition of the direct product   $\bf "10" \times {\bf 8}$. 
Cohen and Lebed obtained analytic expressions of the isoscalar factors as a function of $N_c$ needed for large $N_c$ 
baryon-meson coupling.
Our table was obtained as a direct application of Hecht's Table 4, 
part of which is reproduced in  Table
\ref{Hechtp31} of the Appendix, referring to the irrep ${\bf "10"}$ with multiplicity 2, 
denoted here by  ${\bf "10"}_a$ and ${\bf "10"}_s$ respectively.
For completeness,
to the three rows listed by Cohen and Lebed we have added a fourth one, corresponding to $Y_1$ = $N_c/3-1$, $I_1$ = 2 
and $Y_2$ = 1, $I_2$ = 1/2,
in order to check the orthogonality of columns, given by Eq. (\ref{ORTHOG}),
valid at every $N_c$.  Column ${\bf "35"}$ has  the same phase for all entries as that
of Cohen and Lebed and column ${\bf "27"}$ has opposite phase for all entries.
It may happen that the phase conventions of de Swart and Hecht coincide sometimes. The column ${\bf "10"}_a \equiv {\bf "10"}_2$ 
is entirely different, inasmuch as we use the definition (\ref{REDUCED}) of Hecht to define the representations with multiplicity 2.
We have also added the column ${\bf "10"}_s \equiv {\bf "10"}_1$ where the first three entries vanish at $N_c$ = 3, as observed
in Ref. \cite{Cohen:2004ki}, but the last entry does not. Such a result may be important for 
large $N_c$ baryon studies \cite{Matagne:2014lla}.

In large $N_c$ studies the observables are described by operators expressed  
in terms of SU(6) generators when one considers three flavours, ${N_f}$ = 3. 
The SU(6) generators are components of an irreducible SU(6) tensor operator 
which span the invariant subspace of the adjoint representation denoted here
by the partition $[21^4]$, or  otherwise by its dimensional notation $\bf 35$.
Like for any other irreducible representation its matrix elements can be expressed 
in terms of a generalized Wigner-Eckart theorem \cite{Matagne:2006xx,Matagne:2008kb} 
which factorizes each matrix element 
into products of Clebsch-Gordan coefficients and a reduced matrix element, like in Eq. \ref{FLAVOR}.
The notation is as follows.
The generic name for every generator  is $E^{ia}$. 
An irrep of SU(6) is denoted by the partition $[f]$. 
Then one can write the matrix element of every SU(6) generator $E^{ia}$ as 
\begin{eqnarray}\label{GEN}
\lefteqn{\langle [f](\lambda' \mu') Y' I' I'_3 S' S'_3 | E^{ia} |
[f](\lambda \mu) Y I I_3 S S_3 \rangle =}\nonumber \\ & & \sqrt{C^{[f]}(\mathrm{SU(6)})} 
   \left(\begin{array}{cc|c}
	    S   &    S^i   & S'   \\
	    S_3  &   S^i_3   & S'_3
  \end{array}\right)
     \left(\begin{array}{cc|c}
	I   &   I^a   & I'   \\
	I_3 &   I^a_{3}   & I'_3
   \end{array}\right)  \nonumber \\
& & \times       \sum_{\rho = 1,2}
 \left(\begin{array}{cc||c}
	(\lambda \mu)    &  (\lambda^a\mu^a)   &   (\lambda' \mu')\\
	Y I   &  Y^a I^a  &  Y' I'
      \end{array}\right)_{\rho}
\left(\begin{array}{cc||c}
	[f]    &  [21^4]   & [f]   \\
	(\lambda \mu) S  &  (\lambda^a\mu^a) S^i  &  (\lambda' \mu') S'
      \end{array}\right)_{\rho} , 
   \end{eqnarray}
where $C^{[f]}(\mathrm{SU(6)})$ is the SU(6)
Casimir operator eigenvalue associated to 
the irreducible representation $[f]$, here being the reduced matrix element, followed by the familiar Clebsch-Gordan
coefficients of SU(2)-spin and SU(2)-isospin. The sum over $\rho$ 
contains products of isoscalar factors of SU(3) and SU(6) respectively.
The label $\rho$ is necessary whenever one has to distinguish  between 
irreps $[f']=[f]$ with multiplicities $m_{[f]}$ larger than one in the Clebsch-Gordan series \cite{Matagne:2011fr}
\begin{equation}\label{CG} 
[f] \times [21^4] = \sum_{[f']} m_{[f']} [f']. 
\end{equation}
The two values for $\rho$ both in SU(6) and SU(3) reflects the 
multiplicity problem already appearing in the direct product of SU(3) irreducible representations,
as discussed in Sec.  \ref{Reminder}. It is clear that one must make the sum over $\rho$ in all
cases. 
    The large $N_c$ behaviour is obtained from the analytic expressions of the isoscalar factors 
of SU(3) and SU(6). 
This behaviour is necessary for finding the most dominant contributions in the $1/N_c$ expansion.
Examples of physical interest in baryon spectroscopy for the analytic expressions of SU(6) isoscalar factors 
can be found in
Ref. \cite{Matagne:2014lla} for $[f]$ = $[N_c]$ and $[f]$ = $[N_c-1,1]$. Here we discuss the 
large $N_c$ behaviour resulting from SU(3) isoscalar factors. 

For a comparison with Cohen and Lebed \cite{Cohen:2004ki}
let us consider the column 
${\bf "10"}_a$ of Table \ref{CL1} alone because the column  ${\bf "10"}_s$ is missing in Ref. \cite{Cohen:2004ki}. 
For the first three rows our isoscalar factors are of order $\mathcal{O}(N^{-1}_c)$,   
 $\mathcal{O}(N^{0}_c)$ and $\mathcal{O}(N^{-1/2}_c)$ respectively while from  Ref. \cite{Cohen:2004ki}
Table II at $Y = N_c/3$, $I$ = 1/2, column  ${\bf "10"}_a$ we obtain $\mathcal{O}(N^{0}_c)$,
$\mathcal{O}(N^{0}_c)$ and $\mathcal{O}(N^{-1/2}_c)$ respectively. Thus the large 
$N_c$ behaviour is different from ours for $I_1 = 3/2, Y_1 = N_c/3$, $I_2 = 1, Y_2 = N_c/3$
and $I_1 = 3/2, Y_1 = N_c/3$, $I_2 = 0, Y_2 = 0$. For a proper analysis at large $N_c$ the missing 
column $\rho$ = 2 equivalent to  ${\bf "10"}_s$, is necessary as required by Eq. (\ref{GEN}),
even if some isoscalar factors  vanish  at $N_c$ = 3. 
By summing up the 
contributions from ${\bf "10"}_a$ and ${\bf "10"}_s$ one would expect a similar answer in any convention,
provided the SU(6) isoscalar factors are calculated consistently with those of SU(3). 
Moreover the case $I_1 = 2, Y_1 = N_c/3 - 1$, $I_2 = 1/2, Y_2 = 1$ is missing in Table II
of  Ref. \cite{Cohen:2004ki}, at $Y = N_c/3$ $I$ = 1/2.
Therefore, the results of Ref. \cite{Cohen:2004ki} should be completed with extra rows and columns,
whenever necessary, if one wishes to recover a proper large $N_c$ behaviour. 
In the physical world of $N_c$ = 3 they are sufficient for the exhibited  $I_1, Y_1 $, $I_2 , Y_2$ cases.

It would be interesting to consider further applications of Hecht's SU(3) isoscalar factors 
either in large $N_c$ QCD or in  nuclear physics.


\appendix

\section{}


\begin{sidewaystable}
\caption{Isoscalar factors   
$<(\lambda\mu)Y_1 I_1;(11)Y_2 I_2||(\lambda\mu)Y I>$} of  Hecht's Table 4, p.31 \cite{HECHT}
with corrections for the row $Y_2 = 1$, $I_2 = 1/2$ , $I_1=I+1/2$.
\[
\renewcommand{\arraystretch}{1.7}
\begin{array}{cccc}
\hline
 Y_2 I_2  & \ I_1 &\ \ \ \  (\lambda'\mu') =  (\lambda\mu) & \  (\lambda'\mu') = (\lambda\mu) \\
          &     &\ \ \ \  {\rho} = 1 & \  {\rho} = 2\\ 
\hline
-1 \frac{1}{2} & \ I+1/2 & \ \left[\frac{3(p+1)(\lambda-p)(\mu+2+p)}{2g_{\lambda\mu}(\mu+p-q+1)}\right]^{1/2} 
&  \ \frac{[2g_{\lambda\mu}q-\mu(\lambda+\mu+1)(\lambda+2\mu+6)][(p+1)(\lambda-p)(\mu+2+p)]^{1/2}}
{[\lambda(\lambda+2)\mu(\mu+2)(\lambda+\mu+1)
(\lambda+\mu+3) 2g_{\lambda\mu}(\mu+p-q+1)]^{1/2}}\\

-1 \frac{1}{2} & \ I-1/2 & \ \left[\frac{3(q+1)(\mu-q)(\lambda+\mu+1-q)}{2g_{\lambda\mu}(\mu+p-q+1)}\right]^{1/2} 
&  \ \frac{[2g_{\lambda\mu}p+\lambda(\mu+2)(\lambda-\mu+3)][(q+1)(\mu-q)(\lambda+\mu+1-q)]^{1/2}}
{[\lambda(\lambda+2)\mu(\mu+2)(\lambda+\mu+1)
(\lambda+\mu+3) 2g_{\lambda\mu}(\mu+p-q+1)]^{1/2}}\\

0 0 & \ I & \ - \frac{2\lambda+\mu-3p-3q}{[4 g_{\lambda\mu}]^{1/2} }
&  \ \frac{\sqrt{3}}{2} \frac{\lambda\mu(\mu+2)(\lambda+\mu+1)-\mu(\lambda+\mu+1)(\lambda+2\mu+6)p+\lambda(\mu+2)(\lambda-\mu+3)q
+   2g_{\lambda\mu}pq}
{[\lambda(\lambda+2)\mu(\mu+2)(\lambda+\mu+1)
(\lambda+\mu+3) g_{\lambda\mu}]^{1/2}}\\

0 1 & \ I+1 & 0
&  \ \frac{[2(p+1)(\lambda-p)(\mu+2+p)q(\mu+1-q)(\lambda+\mu+2-q)g_{\lambda\mu}]^{1/2}}
{[\lambda(\lambda+2)\mu(\mu+2)(\lambda+\mu+1)
(\lambda+\mu+3) (\mu+p-q+1)(\mu+p-q+2)]^{1/2}}\\

0 1 & \ I-1 & 0
&  \ - \frac{[2p(\lambda+1-p)(\mu+1+p)(q+1)(\mu-q)(\lambda+\mu+1-q)g_{\lambda\mu}]^{1/2}}
{[\lambda(\lambda+2)\mu(\mu+2)(\lambda+\mu+1)
(\lambda+\mu+3) (\mu+p-q+1)(\mu+p-q)]^{1/2}}\\

0 1 & \ I & \  \frac{[3(\mu+p-q)(\mu+p-q+2)]^{1/2}}{[4 g_{\lambda\mu}]^{1/2} }
& \frac{E}
{2[\lambda(\lambda+2)\mu(\mu+2)(\lambda+\mu+1)
(\lambda+\mu+3)g_{\lambda\mu} (\mu+p-q)(\mu+p-q+2)]^{1/2}}\\

1 \frac{1}{2} & \ I+1/2  & \ \left[\frac{3q(\mu+1-q)(\lambda+\mu+2-q)}{2g_{\lambda\mu}(\mu+p-q+1)}\right]^{1/2} 
    & \ \frac{[2g_{\lambda\mu}p+\lambda(\mu+2)(\lambda-\mu+3)][q(\mu+1-q)(\lambda+\mu+2-q)]^{1/2}}{[\lambda(\lambda+2)\mu(\mu+2)(\lambda+\mu+1)
(\lambda+\mu+3) 2g_{\lambda\mu}(\mu+p-q+1)]^{1/2}}\\

1 \frac{1}{2} & \ I-1/2  & \ - \left[\frac{3p(\lambda+1-p)(\mu+1+p)}{2g_{\lambda\mu}(\mu+p-q+1)}\right]^{1/2} 
    & \ - \frac{[2g_{\lambda\mu}q-\mu(\lambda+\mu+1)(\lambda+2\mu+6)][p(\lambda+1-p)(\mu+1+p)]^{1/2}}
{[\lambda(\lambda+2)\mu(\mu+2)(\lambda+\mu+1)
(\lambda+\mu+3) 2g_{\lambda\mu}(\mu+p-q+1)]^{1/2}}\\[0.5ex]
\hline
\end{array}
\] 
\label{Hechtp31}
\end{sidewaystable}

\begin{table}
\caption{Values of $\lambda'$, $\mu'$, $p$ and $q$ needed for $Y$ = $N_c/3$, $I$ = 3/2 
to calculate the isoscalar factors of  $\bf "10" \times {\bf 8}$ 
using Table  \ref{Hechtp31}. 
The label $(\lambda'\mu')$ identifies the irreps of the
Clebsch-Gordan series (\ref{PROD}) for a given $(\lambda\mu)$ 
in the left hand side.  The isoscalar factors are presented in Table \ref{CL1}.}
\label{param}
\[
\renewcommand{\arraystretch}{1.5}
\begin{array}{cccccc}
\hline
  & \ \lambda' & \ \ \ \ \mu' \ \ \ \ &  \ \  p & \ \ \ q   & \ \ \ (\lambda'\mu') \\
\hline
 ${\bf "35"}$   &  4 &  \ \ \frac{N_c-1}{2} &  \ \   3     &  \ \  \frac{N_c-1}{2}  &  \ \ (\lambda+1,\mu+1)\\
 ${\bf "27"}$   &  2 &  \ \ \frac{N_c+1}{2} &  \ \   2     &  \ \  \frac{N_c-1}{2}  &  \ \ (\lambda-1,\mu+2) \\
 ${\bf "10"}$   &  3 &  \ \ \frac{N_c-3}{2} &  \ \   3     &  \ \  \frac{N_c-3}{2}  &  \ \ (\lambda\mu)\\[0.5ex]
\hline
\end{array}
\] 
\end{table}

In Table \ref{Hechtp31} we  reproduce   part of Table 4 of Hecht's paper  \cite{HECHT} 
which contains the analytic expressions of the isoscalar factors $\langle(\lambda\mu)Y_1 I_1;(11)Y_2 I_2||(\lambda\mu)Y I \rangle$,
often used in quark physics. Note that the entry in the column $\rho$ = 2 for  $Y_2 = 1, I_2 = \frac{1}{2}, I_1 =  I+1/2$  
contains a misprint in Hecht's paper which has been here corrected. This means that in the numerator the bracket $(\lambda+\mu+2-q+1)$
has been  replaced by  $(\lambda+\mu+2-q)$ and in the denominator the bracket $(\mu+p-q)$ has been replaced by $(\mu+p-q+1)$.
In Table  \ref{Hechtp31} we have used $g_{\lambda\mu}$ defined by Eq. (\ref{CSU3}) and $E$ defined by
\begin{eqnarray}
E = \lambda(\lambda+\mu+1)\mu(\mu+2)(2\lambda+\mu+6)+2(\lambda+\mu+1)\mu \nonumber \\
\times [\lambda(\lambda+2)-(\mu+2)(\mu+3)]p -\mu(\lambda+\mu+1)(\lambda+2\mu+6)p^2  \nonumber\\
-2\lambda[(\mu+1)(\lambda+\mu+1)(2\lambda+\mu+6)-\mu g_{\lambda\mu}]q+\lambda(\mu+2)(\lambda-\mu+3)q^2 \nonumber\\
-2[\lambda(\lambda+\mu+1)(2\lambda+\mu+6)-g_{\lambda\mu}]pq+2g_{\lambda\mu}(p^2q+pq^2).
\end{eqnarray}

Table \ref{Hechtp31} and the rest Table 4 of Hecht can straightforwardly be applied 
to a given $(\lambda'\mu')$ with definite values of $Y$ and $I$, from which  one can obtain the integers $p$ and $q$  defined as  
\begin{equation}\label{pq}
Y = p+q - \frac{2\lambda'+\mu'}{2}, \ \ I = \frac{\mu'+p-q}{2}
\end{equation}
given by Hecht where $Y$ is related to the a quantity called $\epsilon$ by
 \begin{equation}
\epsilon = - 3 Y.
\end{equation}
For $\lambda$ = 3 and $\mu = \frac{N_c-3}{2}$ the values of $\lambda'$, $\mu'$ 
together with $p$ and $q$ defined by Eqs. (\ref{pq}) are listed in Table \ref{param}.

We believe there is no reason to
reproduce the full Table 4 of Hecht which contains four distinct tables.

\centerline{\bf Acknowledgments}

This research was supported by the Fond de la Recherche Scientifique - FNRS, Belgium,  under the
grant 4.4501.05.
 

\end{document}